# Optoelectronically Enabled Simultaneous and Decoupled Navigation and Selective Pairing of Electrically Driven Janus Particles for Multi-Agent Microrobotics

Srikanta Debata[1] , Ido Rachbuch[1], Gilad Yossifon[1,2*]

[1]School of Mechanical Engineering, Tel-Aviv University, Tel-Aviv 69978, Israel

[2]School of Biomedical Engineering, Tel-Aviv University, Tel-Aviv 69978, Israel

[*]Corresponding author: gyossifon@tauex.tau.ac.il

## Abstract

Self-propelled Janus particles are promising active agents for microscale manipulation and microrobotic applications. However, when powered by a global external energy source, such as an electric, magnetic, or optical field, all particles respond simultaneously, preventing their independent and concurrent control. Achieving parallel, decoupled control without continuous particle-specific feedback, while preserving intrinsic self-propulsion, remains a significant challenge. Here, we present an optoelectronic (OE) strategy for simultaneously and independently controlling electrically driven, self-propelled metallo–dielectric Janus particles (JPs) using spatially programmable virtual electrodes. These electrodes are generated by projecting a single predefined illumination pattern from a digital micromirror device onto a photoconductive surface. The resulting confinement landscape enables multiple JPs to follow distinct trajectories without continuous monitoring, feedback, or pattern reconfiguration. We demonstrate controlled pairing of JPs with passive polystyrene particles and biological cells. For automated multi-agent operation, an image-based assignment framework identifies optimal agent–target pairs, excludes intersecting trajectories, and generates a single composite pattern for simultaneous, selective pair formation without real-time updating. Finally, we demonstrate independent manipulation of multiple magnetically rolling JPs under a globally applied magnetic field. Overall, this programmable open-loop strategy provides a versatile framework for multi-agent microrobotic manipulation, with potential applications in in-vitro micromanipulation and microfluidic systems.

## 1. Introduction

Active particles at the microscopic scale have emerged as promising tools for targeted delivery, biosensing, microscale assembly, and environmental remediation because of their ability to convert external or locally available energy into directed motion [1–6]. Among the various active colloidal systems, metallo–dielectric Janus particles (JPs) are particularly attractive owing to their structurally simple yet functionally asymmetric design. When subjected to a uniform alternating-current (AC) electric field, these particles can exhibit induced-charge electrophoresis (ICEP) or self-dielectrophoresis (sDEP), depending on the applied frequency and the electrical properties of the surrounding medium [7–11]. Their metallic hemisphere can additionally serve as a mobile microelectrode, promoting localized electrokinetic interactions with passive particles and biological objects and thereby enabling cargo transport, particle assembly, and cellular manipulation [12–18]. Incorporation of a magnetically responsive metallic layer further allows magnetic steering and rolling, providing complementary modes of propulsion and navigation [19–22]. These capabilities make JPs attractive as active microrobotic agents capable of combining locomotion, cargo interaction, and externally controlled functionality [23].

Despite these advantages, the independent and decoupled motion control of multiple active particles remains a major challenge. Electrically and magnetically driven JPs generally respond to fields and gradients which are considered spatially uniform across the workspace, causing all particles within the actuation region to experience the same global input and undergo simultaneous and coupled actuation. Different strategies have been developed to overcome this limitation. Topographical features and physical boundaries, for example, can guide chemically active particles along predefined directions by constraining their accessible motion [24,25]. However, such approaches rely on permanently fabricated structures and consequently offer limited reconfigurability. Optical tweezers provide substantially greater flexibility by using optical gradient forces to trap microscopic objects, while holographic implementations can generate multiple independently controllable traps and transport several particles simultaneously by translating the corresponding optical traps [26,27]. Nevertheless, in these systems, the particle is optically dragged by the moving optical pattern, regardless of whether it possesses self-propelling properties. Moreover, magnetic approaches for the independent manipulation of multiple microrobots are more complex, requiring robots with heterogenous geometries, magnetization states, or other particle-specific responses [28]. Together, these approaches demonstrate the feasibility of multi-particle control while also highlighting the need for a simpler, reconfigurable strategy capable of independently controlling the trajectories of individual particles within a homogenous population.

Optoelectronic tweezers (OET) provide a promising route toward such spatially programmable control. In OET, patterned illumination locally modulates the conductivity of a photoconductive layer and redistributes an externally applied AC electric field to create dynamically reconfigurable virtual electrodes for manipulating microscopic objects through localized dielectrophoretic force [29–34]. Pioneering work by Ming C. Wu group demonstrated the capability of OET for massively parallel manipulation of passive microparticles and cells using projected optical images [35]. When integrated with electrically self-propelled Janus particles (JPs), however, OET can serve a fundamentally different function. Rather than transporting particles solely by moving the projected patterns, electric-field gradients generated at the boundaries between illuminated and nonilluminated regions can locally confine active particles while their intrinsic propulsion drives them along the prescribed paths [36,37]. Building on this principle, previous studies demonstrated optoelectronic reconfiguration of the trajectory of a single JP, enabling its translocation to selected locations and controlled interaction with other JPs [36]. Subsequent work showed second-order directional reversibility of electrically driven JPs in an optoelectronic system [38] and extended the platform to the programmable assembly of multiple active Janus and passive particles [39].

Recent automated OET platforms have used continuous imaging, particle tracking, visual feedback, and real-time path replanning to coordinate multiple particles [40–42]. Although these closed-loop approaches have demonstrated multi-particle control to some extent, they require continuous data acquisition and computation, together with repeated updates of the projected patterns. These limitations highlight the need for a simpler OET-based strategy that can coordinate multiple active particles while preserving their intrinsic self-propulsion and avoiding continuous particle-specific control. Nevertheless, the simultaneous, decoupled, and programmable navigation of multiple active particles under a common global actuation field remains largely unexplored. Automated multi-agent operation further requires the reliable assignment of individual particles to distinct trajectories or targets.

Here, we present a programmable optoelectronic strategy for guiding multiple electrically self-propelled JPs along separate, predefined trajectories encoded within a single projected illumination pattern. The pattern creates multiple virtual-electrode-defined confinement paths, that can be linear or arbitrary shaped. The particles move along these paths through their intrinsic propulsion, enabling concurrent and spatially decoupled navigation without closed-loop tracking, real-time particle-specific steering, or iterative updates of the projected pattern. Programmable optoelectronic confinement is also employed to selectively pair formation between active JPs with passive targets, including passive polystyrene particles and suspended biological cells. To improve scalability and eliminate manual intervention, an algorithm-based framework is integrated into the optoelectronic platform for automated agent-target detection and pairing. Finally, we examine the integration of magnetic rolling with the optoelectronic illumination patterns. In this configuration, a rotating magnetic field provides a global rolling torque, while the optoelectronic confinements selectively restrict particle motion depending on the alignment between the magnetic-field-induced rolling direction and the optical confinement axis, thereby enabling spatially decoupled magnetic actuation and navigation of multiple particles.

## 2. Results and Discussion

### 2.1. Optoelectronic confinement enables parallel and decoupled control of active particles

The motion and control of active Janus particles were investigated using the optoelectronic platform illustrated in Figure 1a, with details of the device fabrication provided in the Experimental Section. Briefly, the system consists of a microfluidic chamber formed between two parallel indium tin oxide (ITO)-coated glass substrates, with the bottom substrate additionally coated with a hydrogenated amorphous silicon (a-Si:H) photoconductive layer. Spatially patterned illumination generated by a digital micromirror device (DMD) locally enhances the conductivity of the photoconductive layer, thereby creating dynamically reconfigurable virtual electrodes with programmable geometries. Under the applied electric field, the JPs exhibit frequency-dependent positive dielectrophoresis (pDEP), coupled with induced electrohydrodynamic effects, resulting in their confinement within the illuminated regions, as illustrated in Figure 1b. By assigning distinct illumination patterns to individual particles, their spatial confinement can be independently programmed, enabling simultaneous and independent control of individual particle trajectories (Figure 1c). To experimentally demonstrate this capability, multiple JPs were initially enclosed within linear illumination patterns oriented in different directions (Figure 1d). Upon application of a 50 kHz, 15 Vpp AC electric field, the particles first underwent electro-orientation, aligning their metallo–dielectric interfaces parallel to the out-of-plane electric-field direction. They subsequently exhibited in-plane motion and reorientation along their respective optical trajectories, with each particle moving independently in a decoupled manner, as demonstrated by the time-resolved trajectories in Figure 1e (Movie S1, SI). Owing to pDEP, the JPs remained confined within the illuminated regions and moved smoothly along the predefined optical trajectories, with their metallic-side-forward self-dielectrophoretic motion [10]. Parallel actuation of the JPs was also examined in the low-frequency regime at 1 kHz and 15 Vpp, as shown in Figure S2a. Under these conditions, the JPs exhibited dielectric-hemisphere-forward induced-charge electrophoretic (ICEP) motion; however, they moved rapidly and became trapped at the optical confinement boundaries. Moreover, passive particles escaped the optoelectronic confinement (Figure S2c, SI), making this frequency unsuitable for the pairing experiments described in the following section. Therefore, 50 kHz and 15 Vpp were selected for subsequent experiments because these conditions enabled smoother, continuous JP motion (also see Figure S2b, SI) while maintaining stable confinement of passive particles for pairing. The ability to independently control the motion of multiple Janus particles was further evaluated using illumination patterns of different geometries (Figure 1f). The particles were simultaneously self-guided along square, triangular, and linear trajectories without requiring external feedback or active control (Movie S1, SI). Their successful motion along these distinct paths demonstrates that the optoelectronic confinement is not restricted to a specific geometry and can be dynamically programmed according to the desired particle trajectory. The versatility of the parallel control approach was also demonstrated using shape-engineered active particles. Plus-shaped particles were fabricated by photolithography [43], followed by selective deposition of a gold layer onto their central lines (Figure 1g). This asymmetric metallic coating introduced the electrical asymmetry required to enable active propulsion. When multiple plus-shaped particles were confined within separate optical trajectories and subjected to a 50 kHz, 15 Vpp AC electric field, they first reoriented such that their central gold-coated lines aligned parallel to the applied field. The particles subsequently moved independently along their respective illumination paths (Figure 1h,i), exhibiting behavior analogous to that of the spherical JPs (Movie S1, SI). These results demonstrate that patterned optoelectronic confinement enables parallel and independent trajectory control of active particles. This approach offers substantial flexibility for controlling multiple active-particle motion in both open-loop [44,45] and closed-loop [46] configurations by dynamically reconfiguring optical patterns without requiring external steering mechanisms or geometrically patterned substrates. Moreover, optically defined virtual trajectories provide a reconfigurable alternative to predefined fabricated electrodes [47,48] or microchannels [46,49], whose fixed geometries require substantial time and resources to fabricate.

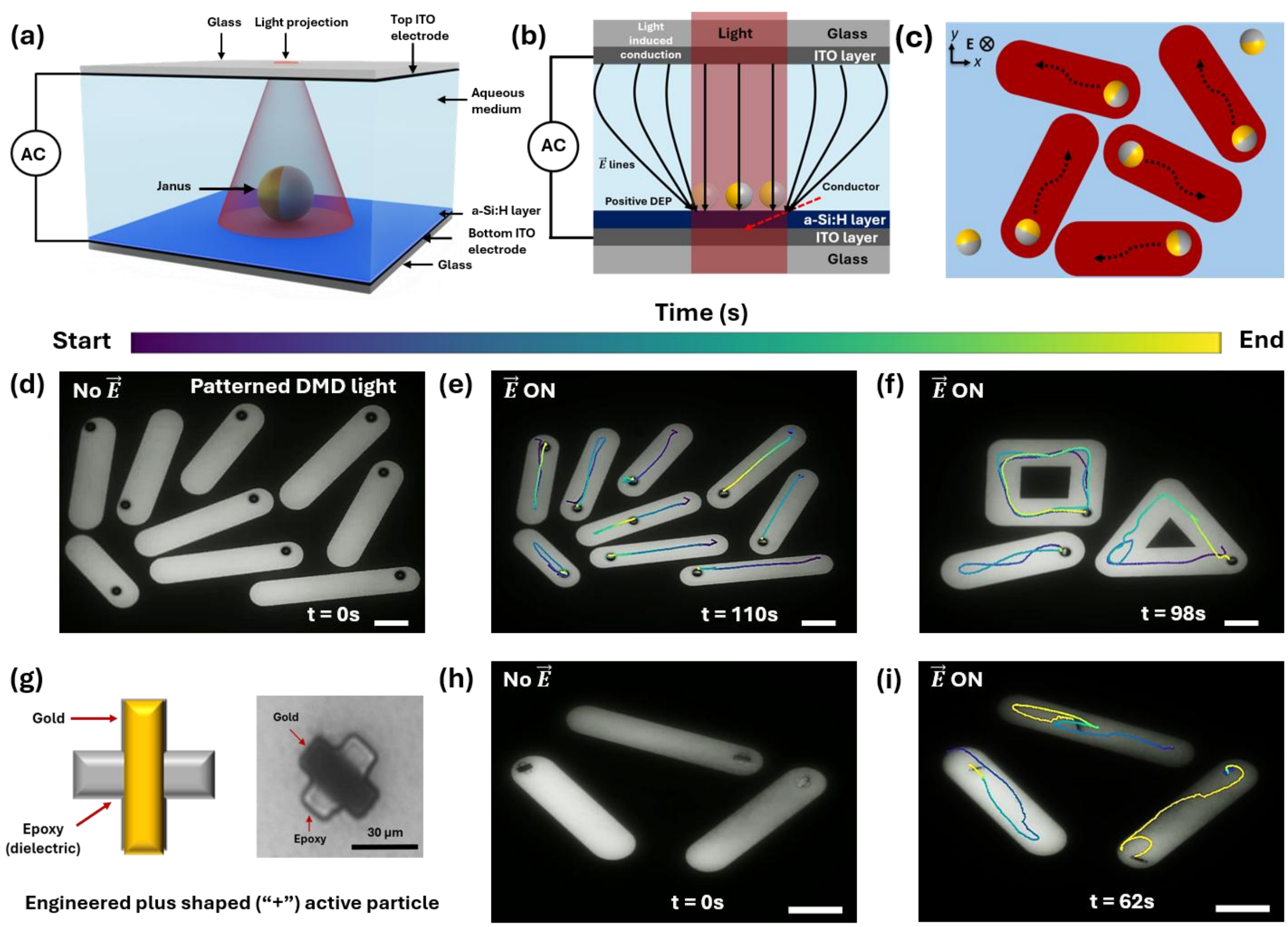

**Figure 1. Parallel and decoupled control of active particles on an optoelectronic platform.** (a) Schematic of the optoelectronic platform. (b) Working principle of the optoelectronic system, in which the illuminated region functions as a virtual electrode and confines the Janus particle (JP) through combined positive dielectrophoresis (pDEP) and inward induced electrohydrodynamics. (c) Schematic illustration of parallel and decoupled control of multiple Janus particles using individually addressable illumination patterns. (d, e) Experimental demonstration of multiple JPs initially confined within separate optical trajectories. Upon application of an AC electric field of 50 kHz and 15 Vpp, the particles aligned with the field direction and subsequently moved along their respective linear trajectories, demonstrating simultaneous and decoupled control. (f) Parallel motion of Janus particles along independently defined square, triangular, and linear trajectories under an AC electric field of 50 kHz and 15 Vpp. (g) Schematic and optical microscopy image of a shape-engineered plus-shaped particle fabricated by photolithography, with its central line

selectively coated with gold. The gold-coated region appears dark in the microscopy image. (h,i) Time-sequence images showing plus-shaped particles confined within separate illuminated trajectories. Following application of an AC electric field of 50 kHz and 15 Vpp, the particles reoriented along the field direction and moved independently within their respective optical regions. Particle trajectories are represented using a time-dependent viridis color map. See Movie S1 in the supporting information for more details. Scale bar: 100 µm.

## 2.2. Selective manual pairing of active particles with passive targets

The ability to independently guide several active particles becomes particularly useful for the selective interaction of active particles with specific targets. Such active particle-target pairing is relevant to active-matter studies, microscale assembly, targeted cargo transport, and potentially selective manipulation and actuation of biological objects. We first demonstrate this concept using manual pairing between active Janus particles and passive polystyrene particles (Figure 2). The schematic in Figure 2a illustrates the sequential steps involved in agent-target pairing. First, the positions of the Janus and polystyrene particles are identified from real-time microscopy images, followed by the manual selection of agent–target pairs. Each selected pair is then individually enclosed within an optical confinement region. Upon application of a 50 kHz, 15 Vpp AC electric field, the JPs migrate toward their corresponding polystyrene targets while remaining confined within the optical boundaries. Upon reaching their targets, the particles attach through induced dipolar attraction [12,13] to form stable pairs. The experimental validation corresponding to the schematic in Figure 2a is shown directly below. To evaluate the role of spatial optical confinement, the controlled pairing process was compared with particle interactions under uniform illumination (Figure 2b). Without optical confinements, the active Janus particles interacted randomly with the passive polystyrene particles, resulting in stochastic and uncontrolled assembly formation (Movie S2, SI). The pairing performance was quantitatively evaluated by comparing the number of targeted particle pairs and their corresponding success rates with and without optoelectronic confinement (Figure 2c). The confined condition exhibited a significantly higher pairing success rate than the unconfined condition, indicating that spatial confinement strongly improves selective agent–target pairing. However, complete pairing was not achieved in some trials due to positional variations, occasional particle escape from the optical regions, and the failure of some active particles to respond effectively to the applied electric field and exhibit self-propulsion.The pairing strategy was subsequently extended from passive microparticles to biological targets. Human K562 cells served as targets (details in ‘Experimental Section’), and each selected JP–cell pair was confined within an optoelectronically defined trajectory (Figure 2d). Upon application of an AC electric field of 50 kHz and 7 Vpp, the JPs migrated toward the respective confined cells and attached to them through positive dielectrophoretic attraction (Movie S3, SI). To demonstrate the potential functionality of the JP–cell attachment, a similar experiment was performed separately using an ITO-based microfluidic device and is presented in Figure S3 (SI). Following attachment, pulsed electric-field stimulation generated localized field enhancement at the metallo–dielectric interface, resulting in selective electroporation of the contacted cell [50]. Electroporation was confirmed by uptake of the surrounding fluorescent dye, while neighboring cells remained unaffected, highlighting the potential of this approach for scalable electroporation-based transfection [50].

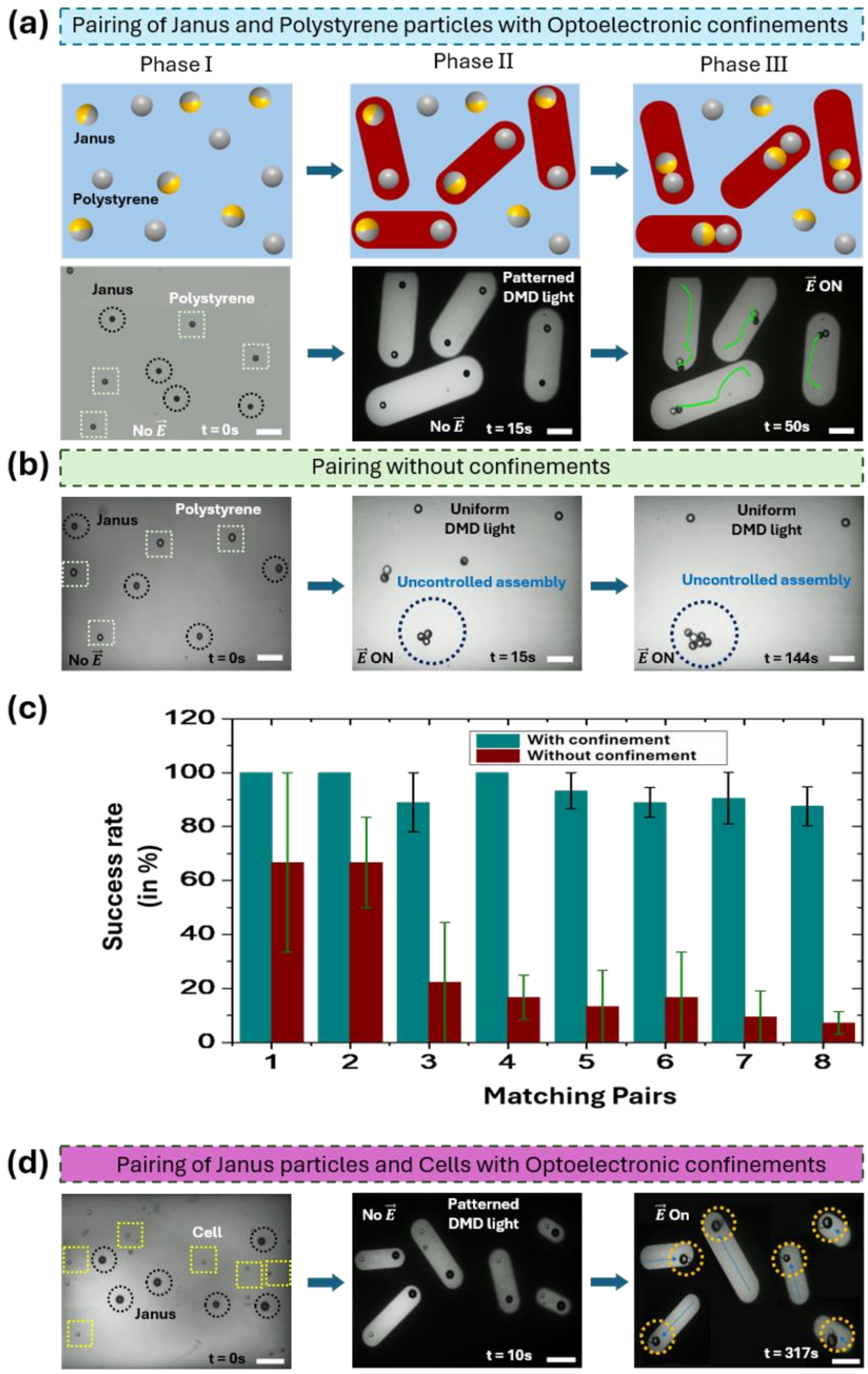

**Figure 2. Manual active–target particles pairing on an optoelectronic platform.** (a) Schematic and time-sequence experimental images showing the different stages of manual pairing between an active Janus particle (Black dashed circle) and a passive polystyrene particle (White dashed square) under an AC electric field of 50 kHz and 15 Vpp. (b)

Janus–polystyrene interactions under uniform illumination, resulting in stochastic and uncontrolled assembly (Blue dashed circle). The AC electric field of 50 kHz and 15 Vpp is applied. See Movie S2 in the supporting information. (c) Graph shows pairing success rate as a function of the number of matched pairs, comparing experiments performed with and without optoelectronic confinement, highlighting the importance of confinements. The experimental observation time was 150s for the pairing. (d) Successful pairing of multiple active Janus particles (Black dashed circle) with targeted biological cells (Yellow dashed square) through optoelectronic confinements and under an AC electric field of 50 kHz and 7 Vpp. See Movie S3 in the supporting information.

## 2.3. Automated Janus particle–Target Pairing for Multiagent Microrobotics

To automate selective pairing and enable scalable multiagent operation, the optoelectronic platform was integrated with an image-based agent-target assignment algorithm, in which active Janus particles (JPs) were designated as agents and passive polystyrene particles as targets. As illustrated schematically in Figure 3a, activation of the automated mode initiated a workflow comprising: (1) live-image acquisition, (2) particle detection and classification, (3) optimal particle-pair assignment and generation of optical confinement patterns via DMD projection, and (4) application of an AC electric field. The detected agents and targets were represented as ($A = \{a_1, a_2 \dots a_N\}$) and ($T = \{t_1, t_2 \dots t_M\}$) respectively, where ($N$) and ($M$) denote the numbers of detected agents and targets. Their two-dimensional ($x$, $y$) center coordinates were used to construct an ($N \times M$) cost matrix containing the Euclidean distance between every possible agent–target combination. The Hungarian algorithm [51–53] then selected up to ($\min(N, M)$) unique one-to-one pairs by minimizing the total assignment distance (Figure S5, SI), thereby preventing duplicate target allocation and providing a globally optimized solution rather than independent nearest-neighbor assignments [54,55] (Figure S6, SI). For each assigned pair, the algorithm generated a linear optical confinement connecting the agent and target centers, with its width set to ($6r_a$), where ($r_a$) is the detected agent radius. This width was selected based on our experimental observations, which showed that the JPs exhibited stable, directed linear motion within an optical confinement of this minimum dimension (Figure S4, SI). Larger widths may result in a less-directed motion toward the agents' targets and reduce the total area available for non-intersecting confinement patterns. The heuristic simulation in Figure 3b illustrates the assignment process, with JP agents shown as blue-green spheres and passive targets as yellow spheres. Following the initial Hungarian algorithm assignment, the proposed trajectories were evaluated for intersections or overlaps. Feasible assignments are shown as white lines, while conflicting trajectories are shown in red and excluded. Confinement regions (yellow) were then generated for the feasible assignments, guiding the simulated agents toward their respective targets for pairing (Movie S5, SI). Following validation of the assignment strategy through heuristic simulation, the complete automated experimental sequence was demonstrated, as shown in Figure 3c. Upon activation of the automated mode, a live microscopy image of the particle distribution was acquired and transferred to the image-processing program. The active JPs and passive polystyrene particles were then detected and classified as agents and targets, respectively, based on their distinct optical characteristics, using the criteria described in the Experimental Section. Each detected particle was subsequently labelled, and the optimized one-to-one assignments were displayed using complementary identifiers, such as ($A_1$) – ($T_1$), ($A_2$) – ($T_2$), and so forth. In the examples presented in Figure S7 (SI), particles without a complementary agent or target, as well as those whose pairing confinement overlapped or intersected with that of another pair, were designated as unpaired (A*/T*) and subsequently excluded from optical projection. Consequently, no confinement patterns were generated for these excluded particles, while each accepted target was uniquely assigned to a single agent. Following assignment, the center coordinates of each agent–target pair were converted into linear optical confinement paths connecting their initial positions. The generated paths were then transmitted as a single input image to the DMD, which projected the corresponding illumination pattern onto the photoconductive substrate. The simultaneous projection of multiple optical confinements enables spatial isolation of the assigned pairs and restriction of each active particle to its designated trajectory. Importantly, the entire automation process was completed in a single step, without continuous real-time monitoring or trajectory updates. Based on the acquired real-time image, the algorithm detected and assigned the agents and targets and generated a single static optical pattern containing all feasible confinement paths, which was then projected simultaneously by the DMD. Once projected, no further trajectory updates were required, enabling passive operation during the subsequent pairing process. Following optical confinement, when an AC electric field of 50 kHz and 15 Vpp is applied, the agents are activated, moved toward and attached to their designated polystyrene targets, forming the intended agent target pairs (Movie S4, SI). This demonstrates parallel manipulation without manual selection or predefined individual trajectories and highlighting the applicability of JPs to multiagent microrobotic operations on the optoelectronic platform.

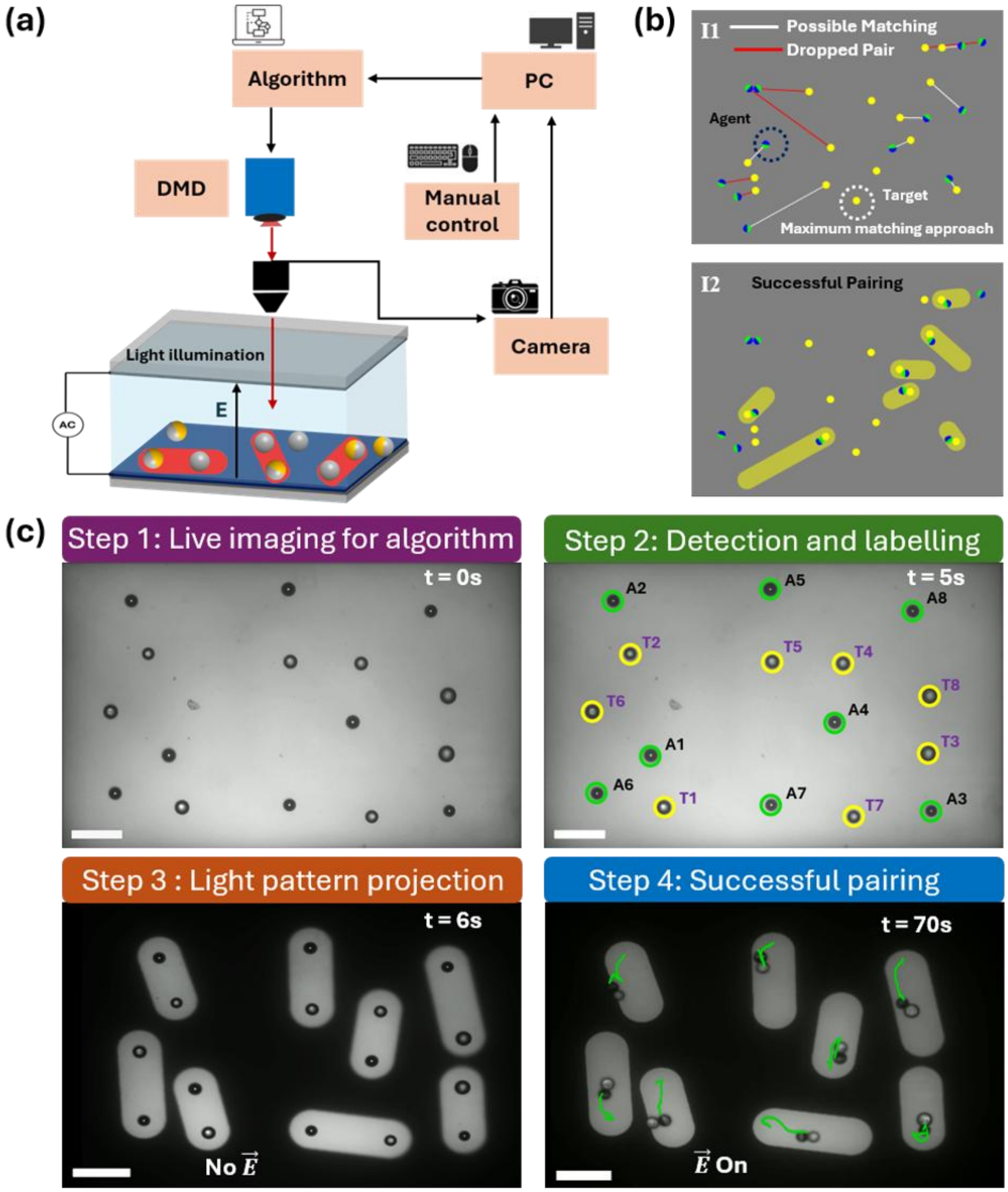

**Figure 3. Algorithm-based automated agent–target pairing.** (a) Schematic of the automated workflow integrating live camera imaging, algorithm-based particle detection, optimal agent–target assignment, DMD-based optical projection and application of an AC electric field. (b) Heuristic simulation showing the optimized agent–target assignments and the subsequent motion of the active Janus particle (JP) agents (blue–green spheres) through the yellow confinement corridors toward the passive targets (red spheres) for pairing. (c) Experimental sequence showing (1) live-image acquisition, (2) particle detection, classification, and complementary labelling (green circles as agents and yellow circles as targets), (3) projection of the corresponding optical confinement paths, and (4) electrically driven motion of the JPs toward their assigned targets and subsequent pair formation. An AC electric field of 50 kHz and 15 Vpp was applied. See Movie S5 in the supporting information. Scale bar: 100 µm.

### 2.4. Decoupling magnetic rolling motion and trajectories of multiple JPs through optoelectronic confinement

To overcome the intrinsically global and nonselective nature of magnetic actuation of identical agents, the optoelectronic platform was combined with magnetic rolling to achieve spatially decoupled control of multiple JPs. In this hybrid actuation mode, the AC voltage was applied continuously to maintain the electric-field gradients and resulting DEP forces at the boundaries of the optical regions, while simultaneously promoting field-induced alignment of the JPs. A rotating magnetic field supplied rolling torque globally to all magnetically responsive particles, whereas the optoelectronic confinement patterns acted as directional gates that determined which particles could undergo net translation. As illustrated schematically in Figure 4a, a particle moved only when its magnetically induced rolling direction was aligned with the corresponding optical confinement path. Although particles confined to trajectories with different orientations experienced the same global magnetic actuation, those whose confinement axes were not aligned with the magnetically induced rolling direction were driven toward the optical boundaries. Unable to cross these boundaries, the particles remained trapped within their respective confinements.

Under a fixed magnetic rotation rate of 30 rpm, when the effective rolling thrust normal to the boundary remained below the maximum DEP counteracting force, a stable force-equilibrium position was established [56–58]. Under these conditions, the particles could continue rotating or undergo local slipping near the boundary without escaping the confinement or exhibiting appreciable net translation. In contrast, when the magnetic rolling direction was aligned with the longitudinal axis of the confinement, the particle could translate through the confinement while the transverse DEP counteracting force prevented lateral escape. Consequently, changing the orientation of the rotating magnet altered the rolling direction and selectively released particles confined along the newly aligned trajectories, while particles in the remaining patterns stayed locked. The experimental sequence in Figure 4b demonstrates this decoupled operation along both horizontal (x-axis) and vertical (y-axis) directions (Movie S6, SI). The recorded color-coded trajectories (Blue for horizontal and red for vertical) confirm that translation occurred only within optical paths aligned with the instantaneous magnetic rolling direction. Using this hybrid actuation strategy, JPs were guided along rectangular and triangular confinement paths by sequentially changing the magnetic rolling direction while continuously keeping the electric field turned on, as shown in Figure S8 (SI) (Movie S7, SI). This approach was further employed for agent–target pairing, as shown in Figure 4c. Active JPs and passive polystyrene targets were confined within predefined optical paths. By adjusting the magnetic rolling direction, the selected JPs were transported along the corresponding confinement toward their designated targets, where they formed the intended pairs (Movie S8, SI). The contribution of optical confinement to pairing under hybrid electric–magnetic actuation was further quantified by statistical analysis (Figure S9, SI), which revealed significantly higher pairing success rate under confined conditions than under unconfined conditions. Here, the unconfined condition refers to global illumination combined with spatially uniform magnetic and electric fields. Under these conditions, magnetic rolling occurs along a single axis without active steering. Even if steering were applied, all particles would respond collectively, making independent navigation and selective pairing difficult to achieve. For biomedical applications, increasing the solution conductivity to physiologically relevant levels weakens the confining DEP force [59,60], which can cause the particles to escape the optoelectronic confinement during magnetic rolling. Extending the platform to higher-conductivity conditions may be achieved by employing phototransistor-based optoelectronic devices, as reported by the Ming C. Wu group [61] or by replacing the uniformly coated photoconductive layer with spatially patterned photoconductive structures [62] to enhance the induced in-plane DEP force.

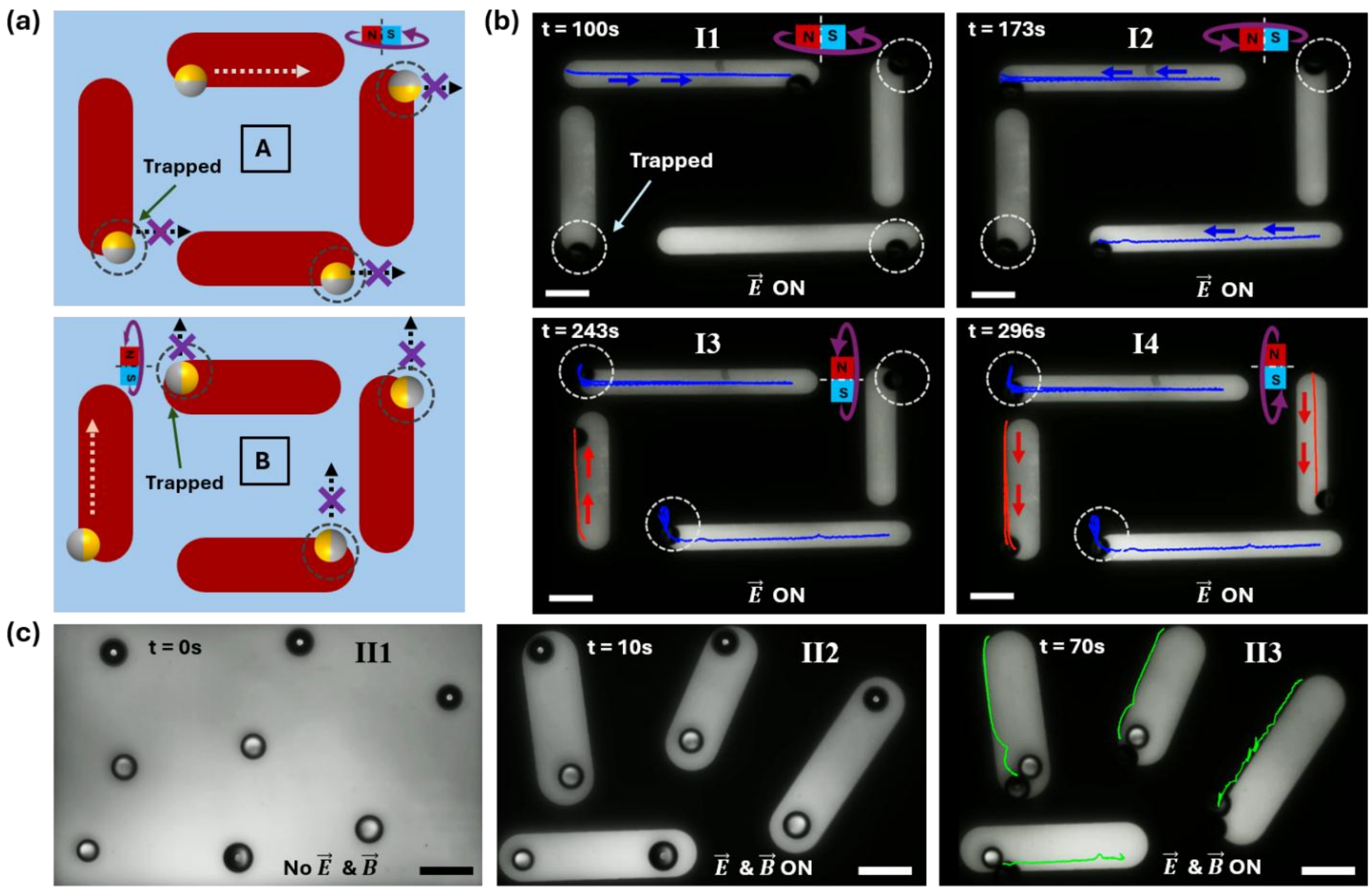

**Figure 4. Decoupling the magnetic rolling motion and trajectories of Janus particles through optoelectronic confinements.** (a) Schematic of the hybrid actuation strategy, in which a global rotating magnetic field drives particle rolling while optoelectronic confinement selectively permits translation only along trajectories aligned with the rolling direction. Particles confined along nonaligned paths remain "trapped" at the optical boundaries due to counteracting DEP force. (b) Experimental demonstration of direction-selective motion along horizontal (I1, I2) and vertical confinement paths (I3,I4). Sequential reversal of the magnetic rolling direction selectively released particles within aligned trajectories, whereas the remaining particles were retained by the opposing DEP force. Color-coded trajectories indicate particle motion, blue: along horizontal and red: along vertical. See Movie S6 in the supporting

information. (c) Hybrid electric–magnetic agent–target pairing, in which the selected Janus particles were rolled along optical confinement paths toward passive polystyrene targets and formed pairing as shown in II1-II3. See Movie S8 in the supporting information. A magnetic rotation rate of 30 rpm and an AC electric field of 5 MHz and 20 Vpp were applied. Solution conductivity: 7 µS/cm. Scale bar: 50 µm.

## 3. Conclusion

In summary, we developed a programmable optoelectronic platform for parallel and spatially decoupled navigation of electrically driven metallo-dielectric Janus particles (JPs) along predefined linear and complex trajectories. A single DMD-generated illumination pattern, combined with an AC electric field, establishes persistent virtual-electrode boundaries that confine the particles through DEP while their intrinsic propulsion drives motion along the designated paths. Unlike conventional closed-loop approaches requiring continuous tracking, particle-specific steering, and repeated pattern updates, the system operates autonomously after the initial projection and field activation. This open-loop strategy reduces continuous intervention and computational demands while providing a simple, scalable approach for controlling larger active-particle populations.

The optoelectronic confinement approach was also used to selectively pair active JPs with passive polystyrene particles or biological cells. Confinement-based pairing achieved a significantly higher success rate than unconfined uniform illumination, which primarily produced stochastic assemblies rather than targeted one-to-one pairs. To automate pairing, an image-based algorithm identified JPs as agents and passive objects as targets, constructed a distance-based cost matrix from their detected positions, and applied the Hungarian algorithm to determine optimal one-to-one assignments. Assignments with intersecting confinement paths were excluded, and all feasible paths were combined into a single illumination pattern and projected simultaneously. This enabled parallel pair formation after a single detection and assignment step, without continuous tracking, trajectory recalculation, or pattern updates.

The platform was further combined with magnetic rolling to overcome the globally coupled nature of magnetic actuation. At a magnetic rotation rate of 30 rpm and under an AC electric field of 5 MHz and 20 Vpp, particles translated only when the magnetic rolling direction aligned with their optical trajectories, while those on nonaligned paths remained confined by the opposing DEP restoring force induced at the illumination boundaries. Sequentially changing the magnetic rolling direction thus selectively activated different particles under the same global magnetic field, with the optical paths acting as spatial gates that converted a uniform magnetic input into trajectory-dependent responses.

Collectively, these findings establish programmable optoelectronic confinement as a versatile framework for spatially decoupled, open-loop multi-agent control, with broad potential for scalable microscale manipulation and selective particle–cell interactions.

Future work could extend the assignment algorithm to consider JP properties and target characteristics beyond distance, including size, speed, cargo capacity, and target state. Image-based features such as cell morphology or fluorescence could enable more selective JP–cell matching. At the device level, patterned photoconductive or phototransistor-based layers could generate localized electric fields in conductive physiological media. Combined with controlled cargo release, these advances could support selective cell treatment, electroporation, and targeted drug delivery.

## 4. Experimental Section

### 4.1. Fabrication of Janus particles, engineered particles and solution preparation

For the fabrication of Janus particles (JPs), commercially available polystyrene (PS) microparticles (Sigma-Aldrich; diameter ~27 μm) were used as the starting material. The concentrated PS particles were washed three times by centrifugation in isopropanol (IPA) and subsequently diluted in the same solvent to obtain an appropriate particle density. The washed particle suspension was then drop-cast onto a thoroughly cleaned glass slide to form a monolayer. The IPA was allowed to evaporate naturally, leaving a monolayer of PS particles on the glass surface. Before metal deposition, the particle-coated glass slides were treated with oxygen plasma for 1 min. Subsequently, 15 nm Cr, 50 nm Ni, and 15 nm Au layers were sequentially deposited onto the particles under high vacuum ($10^{-7}$ mbar) using an electron-beam evaporator (VST TFDS-680), producing the metallo-dielectric Janus particles. The resulting JPs were removed from the glass substrate by gently scratching and stored in deionized (DI) water. The engineered plus ("+") shaped particles are designed through the standard photolithography technique, where the metal (Au) is coated on the one wing of the epoxy body as discussed by our previous study [43]. Before the experiments, the particles were rinsed three times with the prepared conductive solution. The conductive solution consisted of 10 μM KCl and 0.01% (v/v) Tween 20 and had a conductivity of 6–7 μS/cm.

For cell experiments, the human myelogenous leukemia cells K562 was used which is grown through standard cell culture protocols. Briefly, the cells were cultured in RPMI 1640 medium (Biological Industries) supplemented with 10% heat-inactivated fetal bovine serum (FBS) (Biological Industries), 1% penicillin–streptomycin (Biological Industries), and 2% L-glutamine (Biological Industries) and maintained at 37 °C in a humidified incubator containing 5% $CO_2$. Fluorescence labeling was performed by maintaining the cells at 37 °C for 10 min under the same culture conditions in the presence of 5 μg/mL 6-carboxyfluorescein diacetate (CFDA; Sigma-Aldrich) dye. Before the experiments, the cells were thoroughly washed with 300 mM sucrose solution and subsequently resuspended in the same isotonic solution. Propidium iodide (PI; Sigma-Aldrich) was then mixed to a final concentration of 50 μg/mL. The JPs and cells were then mixed in 300 mM sucrose solution immediately before the experiments. For each experiment, 10 μL of the prepared suspension was introduced into the experimental chamber.

## 4.2. Optoelectronic device fabrication

The optoelectronic (OE) microfluidic device was constructed by placing a ~130 μm-thick spacer (Grace-Bio) between two parallel ITO-coated glass substrates, forming the microfluidic chamber shown in Fig. 1a. The ITO electrodes were approximately 200 nm thick (Sigma-Aldrich). To introduce photoconductive functionality into the bottom electrode, a hydrogenated amorphous silicon (a-Si) film with a thickness of approximately 900–1000 nm was deposited onto the ITO-coated glass substrate by plasma-enhanced chemical vapor deposition (PECVD) using a Plasma Lab 100 system (Oxford Plasma Instruments). In the PECVD chamber, $SiH_4$ and $H_2$ were introduced at flow rates of 20 and 60 sccm, respectively, while maintaining a substrate temperature of 200 °C, a chamber pressure of 1000 mTorr, and an RF power of 30 W. The photo response of the deposited a-Si:H film was characterized in our previous studies by comparing its electrical conductivity under illuminated and dark conditions [36,38]. For this characterization, a photoresistor test structure with approximate dimensions of 410 × 500 μm² was designed by photolithography [36]. The test structure was fabricated on a 1-mm-thick conventional glass substrate coated with a-Si using the same PECVD conditions described above. To establish electrical contacts, a 50-nm-thick Ti adhesion layer followed by a 200-nm-thick Au layer was deposited onto the patterned structure using an electron-beam evaporator (VST TFDS-680). The resulting metal contacts were subsequently soldered to electrical wires for conductivity measurements.

## 4.3. Experimental setup and operation

The OE experimental platform was constructed by coupling a DMD-based illumination system (Mightex Polygon 1000-G; 625 nm wavelength; ~1.4 W LED source) to an upright optical microscope (Nikon ECLIPSE Ni-E), as schematically shown in Figure S1 (SI). The microscope was equipped with motorized translation stages (Marzhauser Scan Plus 100 × 85, 4 mm lead-screw pitch), a Prime BSI Express camera, and a Nikon fluorescence LED illumination source. This configuration enabled controlled projection and imaging of spatially patterned light on the photoconductive layer. The DMD and camera acquisition were simultaneously controlled using a custom Python program, enabling projection of either predefined patterns or real-time trajectory drawing on the live image, while also supporting automated pattern generation and projection through algorithm-based image processing. The details about photoelectrical response of the deposited a-Si:H layer were independently evaluated using the fabricated photoresistor structure and mentioned in our previous studies [36,38]. Briefly, the structure was exposed to different intensities of 625 nm illumination generated by the OE system, while the corresponding current–voltage (I–V) characteristics were acquired simultaneously using a Keithley 2636A source meter. At a DC bias of 5 V and a maximum illumination intensity of approximately 0.65 W/cm² applied from the top, the conductivity of the a-Si:H structure under illumination increased by approximately two orders of magnitude relative to its dark-state conductivity. Electrical activation of the OE device was achieved by attaching copper tape (Digi-Key, 3M9887-ND) to the two ITO-coated glass substrates. Sinusoidal AC voltages ranging from 10 to 15 Vpp were generated using an Agilent 33250A function generator at frequencies of either 50 kHz or 5 MHz. The applied electrical signals were monitored using a Tektronix TPS-2024 oscilloscope during the experiments.

### 4.4. Microscopy and image analysis

The motion of the particles inside the OE microfluidic chamber was monitored using a Nikon ECLIPSE Ni-E microscope fitted with a 10× / 20x objective. Image sequences were acquired at 20 frames per second using a Prime BSI Express camera. The recorded videos were subsequently processed using a custom Python script developed for particle tracking and kinematic analysis. Particle trajectories were extracted using the Channel and Spatial Reliability Tracking (CSRT) method implemented in the OpenCV library, from which the instantaneous particle positions and velocities were determined. To facilitate visualization of temporal changes in particle motion, the extracted trajectories were represented using time-dependent color coding.

### 4.5. Algorithm-Based Agent-Target Detection and Matching

An automated image-processing and agent-target matching algorithm was developed in Python using OpenCV, NumPy, and SciPy to identify and pair active Janus particles (agents) with passive polystyrene particles (targets). The acquired images were first smoothed using a Gaussian filter with a 9 × 9-pixel kernel and a standard deviation of 2 pixels ($\sigma = 2$). Candidate particles were detected using the Hough circle transform (with dp=1.2), a minimum center separation of 20 pixels and a detectable radius range of 8-80 pixels. The detected particles were classified as agents or targets based on their grayscale intensity distribution, exploiting the dark metallic region of the Janus particles and the brighter appearance of the translucent polystyrene particles. A grayscale threshold of 110 was used for classification, with particles assigned as agents when the fraction of dark pixels (i.e. under the threshold) within the detected particle exceeded 55% and the normalized radius of the bright central region was below 0.35. Otherwise, they were classified as targets. When circle detection was unsuccessful, a contour-based detection procedure using the same grayscale threshold and particle-size limits was employed as a fallback. Following detection, the Cartesian coordinates of the agents and targets were extracted, and a cost matrix was constructed from the pairwise Euclidean distances between all agent-target combinations. The optimal one-to-one assignments were then determined using the Hungarian algorithm through the linear-sum-assignment method implemented in SciPy, thereby minimizing the total agent-target distance. For each assigned pair, a directed linear confinement trajectory was generated between the agent and target positions, with the confinement width defined as ($6r_a$), where ($r_a$) is the detected agent radius (Figure S4, SI), and a minimum width of 17 pixels. The resulting trajectories were subsequently used to generate the corresponding optoelectronic illumination patterns, allowing automated and simultaneous confinement of the identified agent-target pairs. The decision to project or reject each linear optical confinement was based on the availability of a complementary agent–target pair and whether the proposed confinement intersected or overlapped with that of another pair. Projection was rejected if no complementary pair was available or if the proposed confinement intersected or overlapped with another confinement.

### 4.6. Hybrid optoelectronic-magnetic propulsion

For the hybrid mode of propulsion, the JPs were actuated using a combination of optoelectronics and magnetic fields. Prior to the experiments, the JPs were magnetized along the long-axis of their metallic hemisphere (to support synergistic rolling under the electric field alignment) by placing the particle-containing slide between two permanent neodymium magnets with opposing poles facing each other. Following magnetization, magnetic rolling motion was induced by rotating a neodymium permanent magnet ($14 \times 12 \times 19$ mm$^3$) with a DC motor positioned approximately 5 cm from the center of the microfluidic chamber. This distance was maintained to minimize magnetophoretic displacement of the particles. The orientation of the rotating magnet was adjusted according to the orientation of the optoelectronic confinement trajectories, allowing the magnetic rolling direction to be aligned with or orthogonal to the defined optical pathways.

## Acknowledgments

G.Y. acknowledges support from the Israel Science Foundation (ISF) grant #1429/24. Moreover, we thank the Center for Nanoscience and Nanotechnology, Hebrew University of Jerusalem for assisting us in fabricating the photoconductive substrates. We would like to thank Stav Ashur and Professor Avishai Sintov for their valuable discussions and suggestions regarding the development of the algorithm employed for multiagent applications. We also thank Eyal Abraham and Shlomo Wais for constructing the joysticks controlled magnetic system operated through Arduino Nano and for making the Arduino Nano operated relay system for pulse generation.

## Conflict of Interest

The authors declare no conflict of interest.

**Optoelectronically Enabled Simultaneous and Decoupled Navigation and Selective Pairing of Electrically Driven Janus Particles for Multi-Agent Microrobotics**

Srikanta Debata[1] , Ido Rachbuch[1], Gilad Yossifon[1,2*]

[1]School of Mechanical Engineering, Tel-Aviv University, Tel-Aviv 69978, Israel

[2]School of Biomedical Engineering, Tel-Aviv University, Tel-Aviv 69978, Israel

# Supporting information

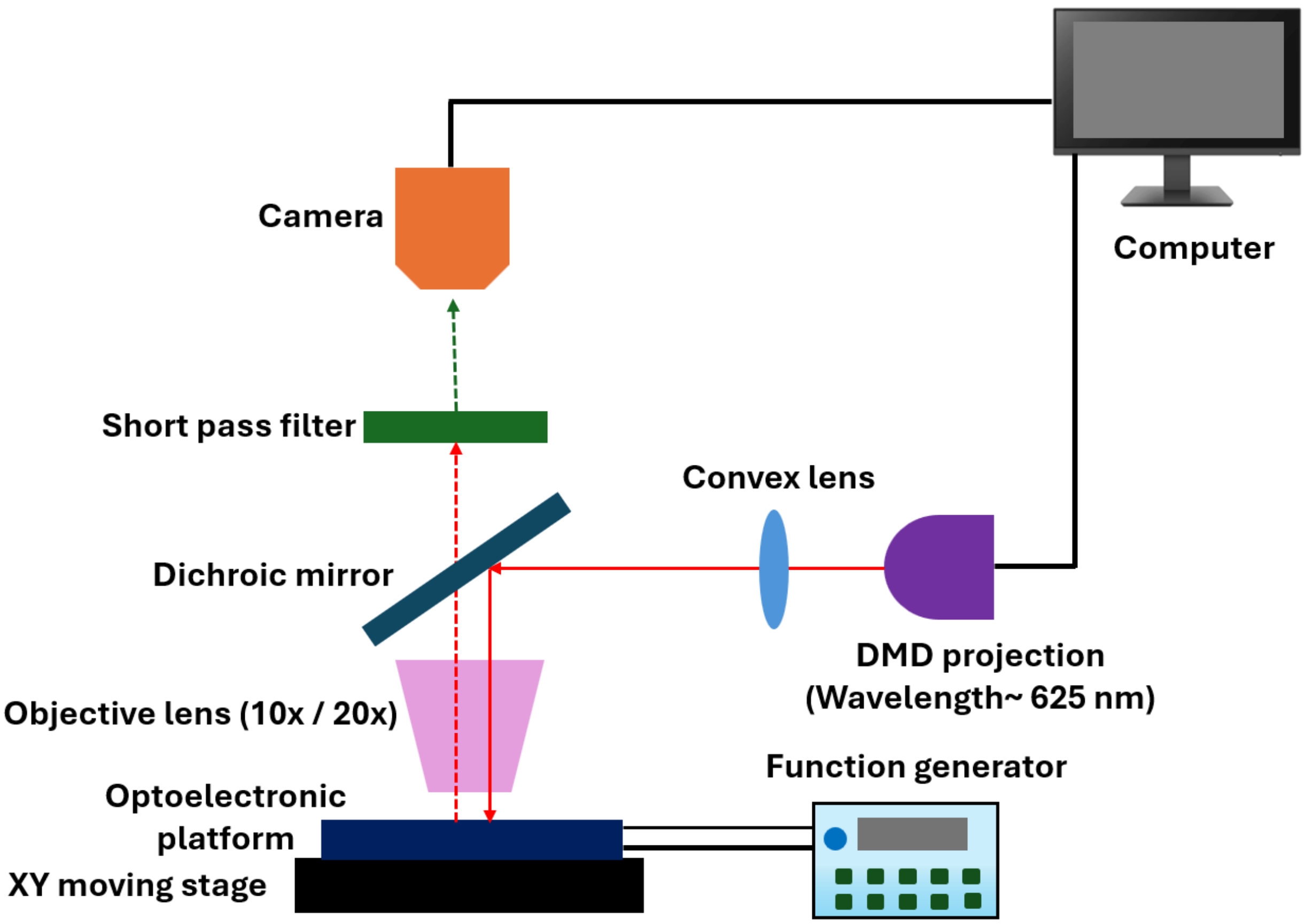

**Figure S1. Schematic of the experimental setup used for our study.**

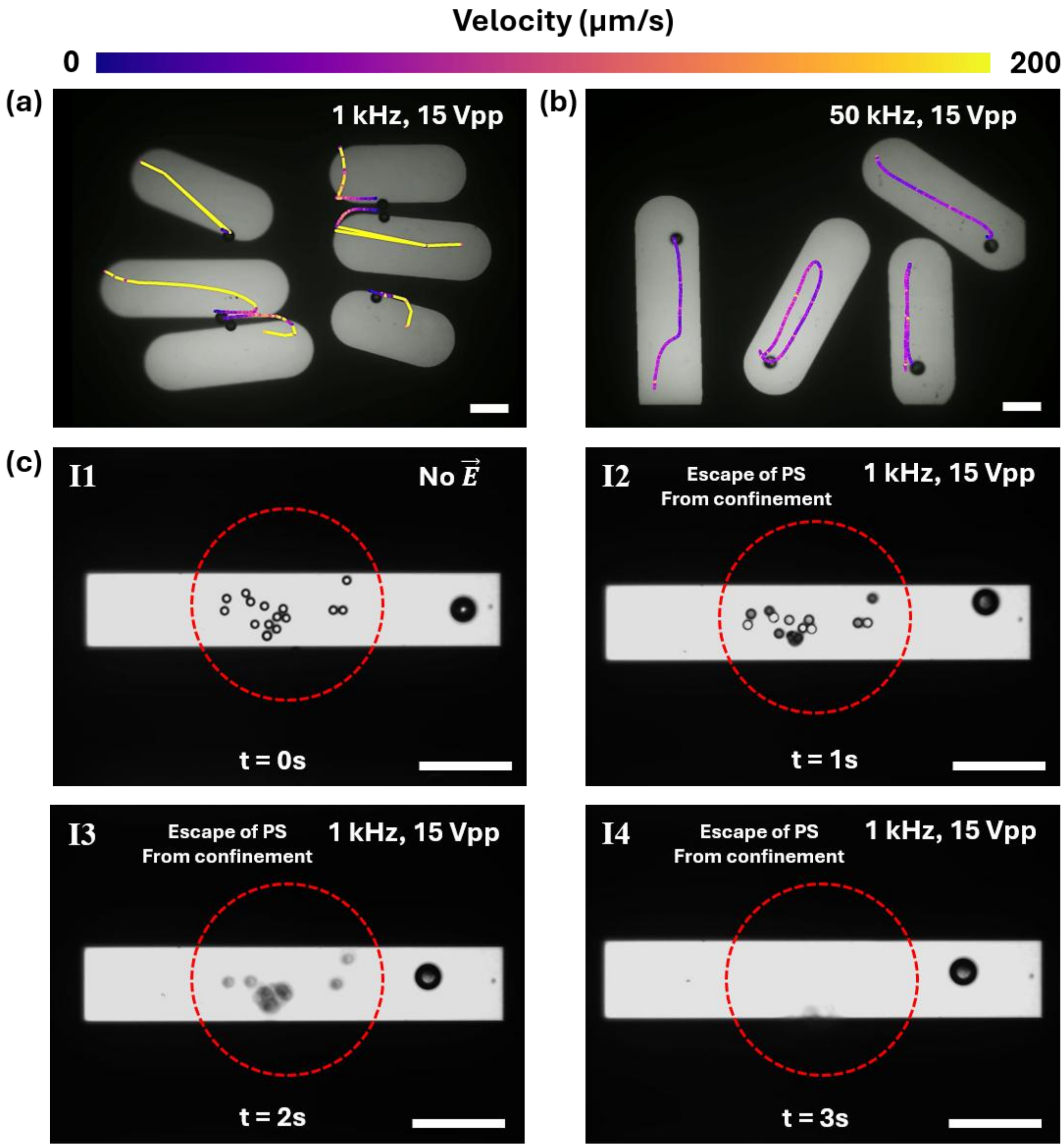

**Figure S2. Immobilization of JPs at the optical-region boundary and escape of polystyrene particles at 1 kHz, compared with the smoother, continuous motion of JPs at 50 kHz.** (a) At 1 kHz and 15 Vpp, induced-charge electrophoresis (ICEP) produced irregular JP motion within the confinements. Upon reaching the confinement boundary, the JPs were immobilized. (b) At 50 kHz and 15 Vpp, self-dielectrophoretic (sDEP) propulsion resulted in smoother and continuous JP motion due to lesser ACEO flow comparative to ICEP with a more uniform speed. (c) I1–I4 show the rapid sequential escape of 10 µm passive polystyrene (PS) particles (outlined by dashed red circles) from the optical confinement immediately after applying a 1 kHz, 15 Vpp electric field. Scale bars: 100 µm. The color of trajectory varies according to the velocity of the particles.

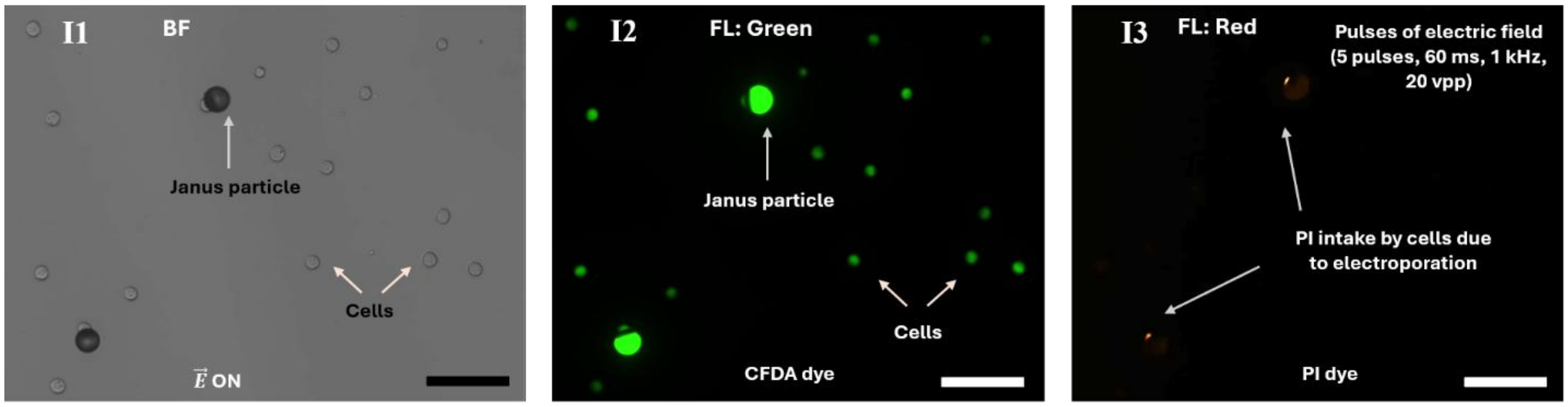

**Figure S3. Localized electroporation of cells on an ITO substrate using Janus particles.** (I1) Bright-field image showing attachment of a Janus particle (JP) to a cell through positive dielectrophoresis, followed by its jumping over the cell. An AC electric field of 50 kHz and 7 Vpp was applied for JP–cell attachment. After attachment, the electric field was adjusted to 1MHz and 10 Vpp to supress the JP motion and enhance cell retention. (I2) Corresponding green-channel fluorescence image showing the CFDA-stained cell and fluorescent polystyrene-based JP, both indicaged by labelled arrows). (I3) Following attachment, a train of five electric pulses at 1 kHz and 20 Vpp, each with a duration of 60 ms and separated by 1 s, was applied to induce electroporation. The resulting red fluorescence corresponds to propidium iodide (PI) uptake from the surrounding medium through the electroporated cell membrane, indicating that cell electroporation occurred specifically in the cell attached to the JP (Movie S4, SI). Scale bars: 100 µm.

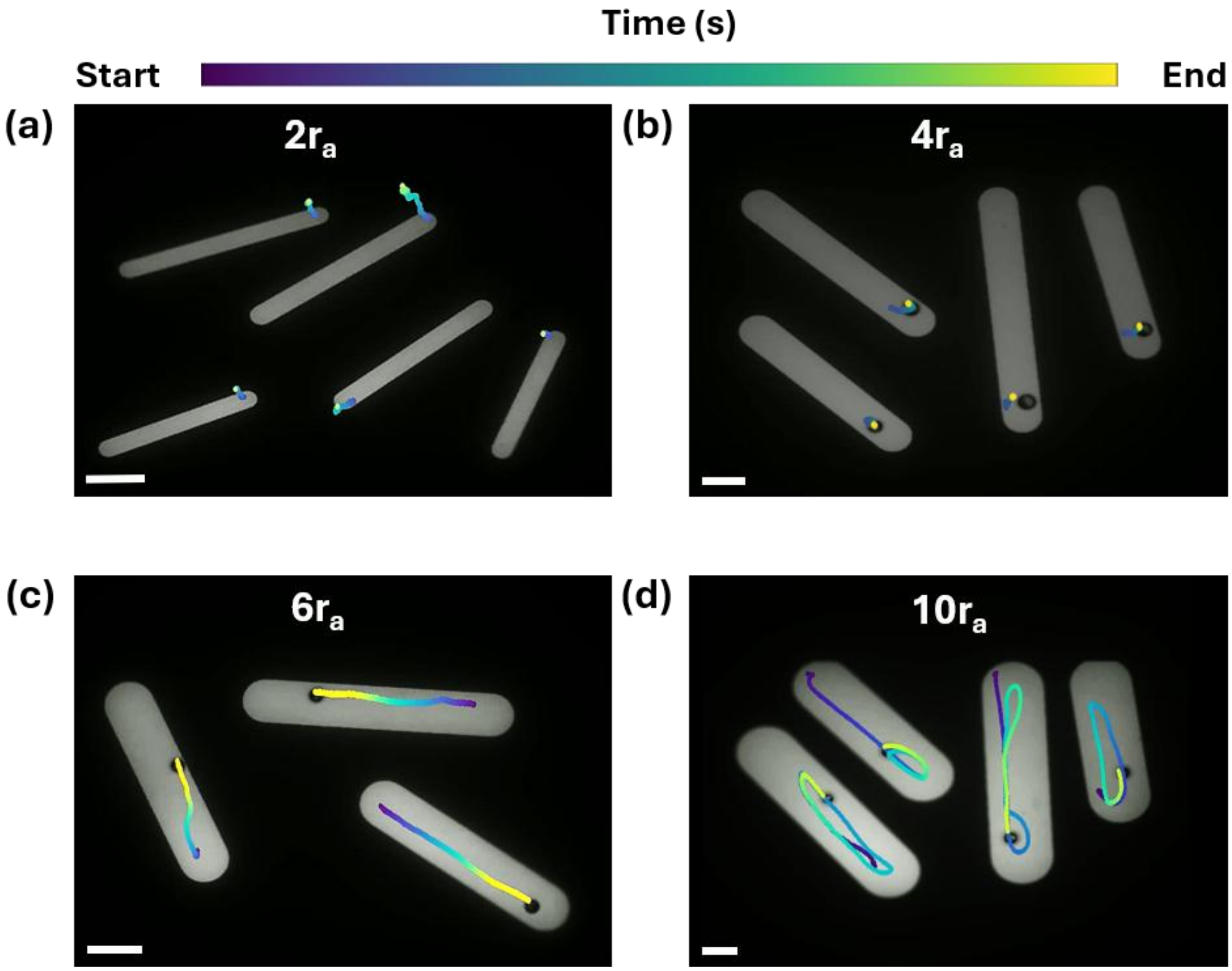

**Figure S4. Effect of optoelectronic confinement width on Janus particle (JP) motion.** (a–d) JP motion within confinement paths having widths of ($2r_a$), ($4r_a$), ($6r_a$), and ($10r_a$), respectively, where ($r_a$) denotes the JP radius. Smooth particle motion was observed when the confinement width reached ($6r_a$). At a width of ($2r_a$), which is equal to the JP diameter, the particles either escaped the confinement or remained near its boundary without appreciable forward motion. The JPs were tracked for 42 s, and their time-dependent trajectories are shown. Applied AC electric field: 50 kHz and 15 Vpp. Scale bar: 100 µm.

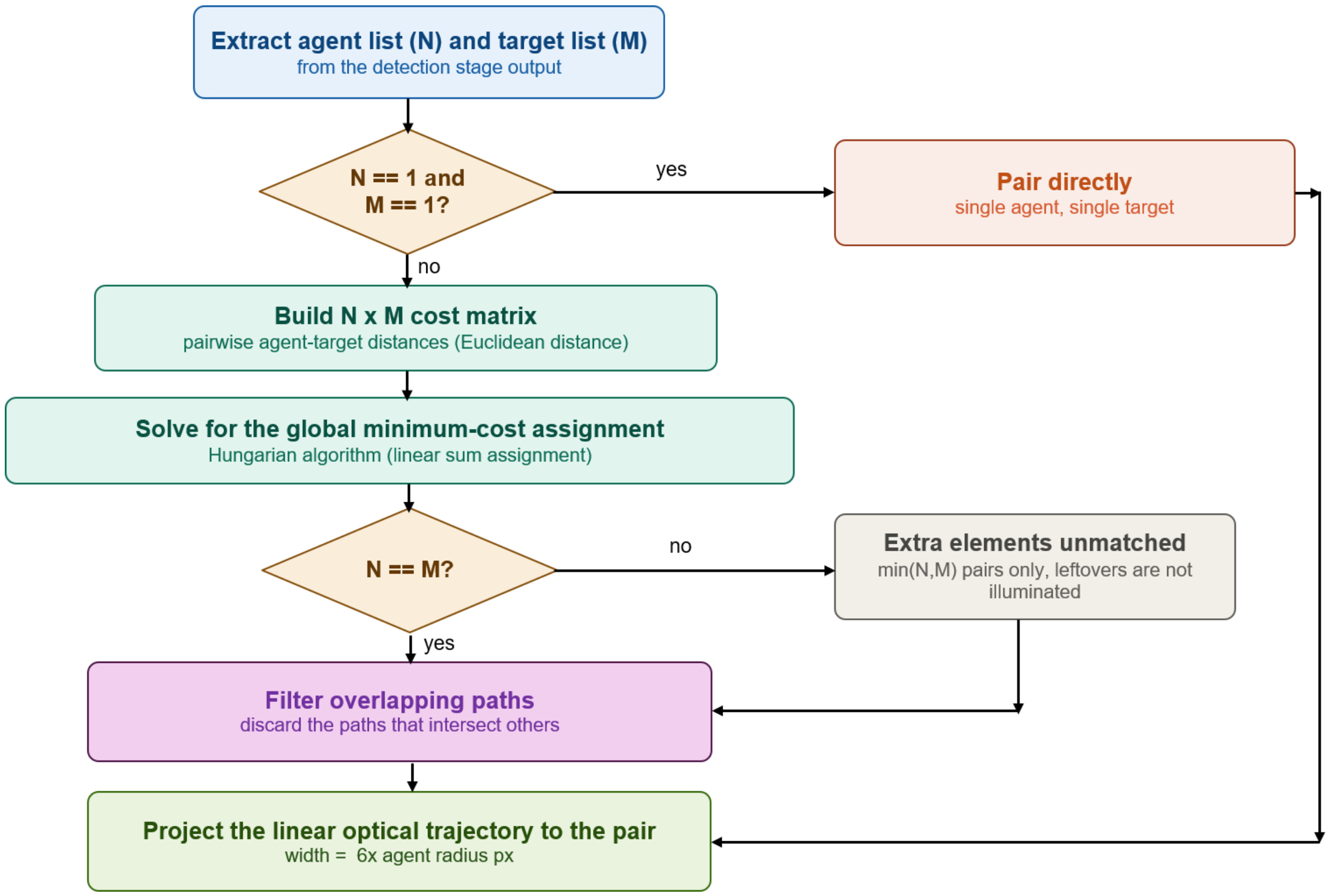

**Figure S5.** Agent-target pairing algorithm logic. The assignment step is a single global optimization: it evaluates every possible pairing combination at once and returns the one with the lowest total distance, rather than each agent greedily picking its own nearest target or real time updating.

| | Target T1 | Target T2 |
|---|---|---|
| **Agent A1** | **50 px** | 120 px |
| **Agent A2** | 90 px | **60 px** |

**Chosen pairing (global minimum)**
A1 to T1, A2 to T2 · total = 110 px

**Alternative pairing (rejected)**
A1 to T2, A2 to T1 · total = 210 px, higher cost

**Figure S6.** Rationale for the selection of specific agent-target pairs. The Hungarian algorithm evaluates every possible pairing at once and returns the one with the lowest total distance.

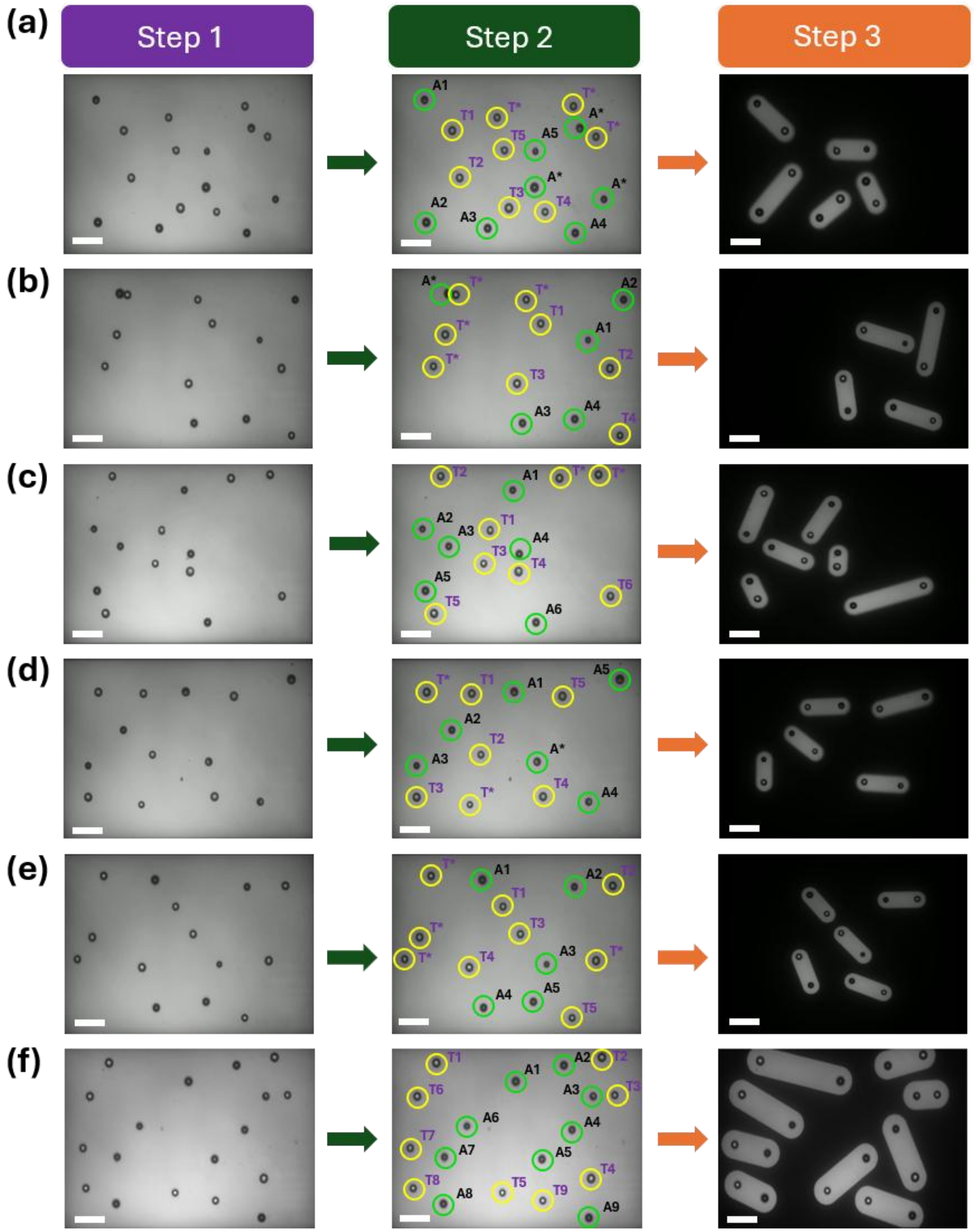

**Figure S7. Verifying the reliability of algorithm.** (a-f) Some examples of algorithm-based detection of agent-target pairs. A*/T* denotes an agent/target that cannot form a pair due to the absence of a complementary partner or intersection with another previously identified pair.

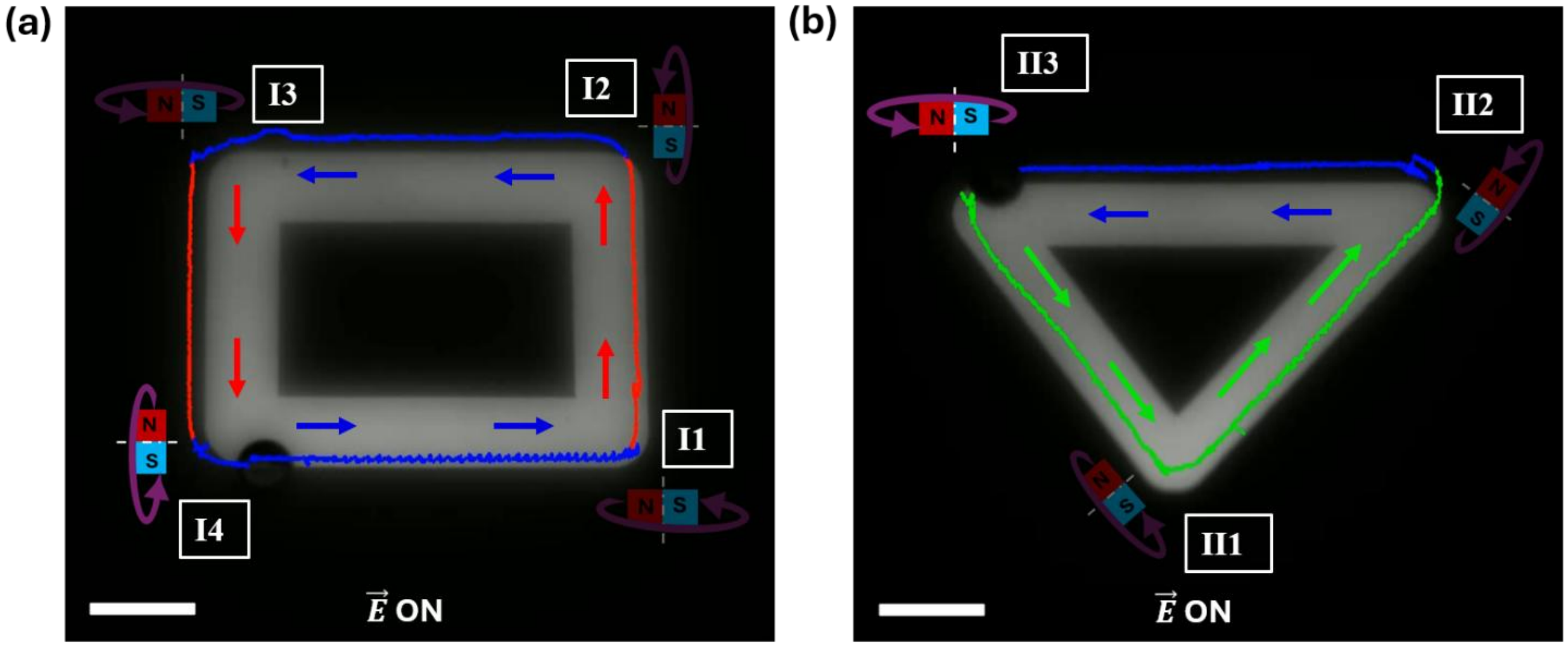

**Figure S8. Sequential motion of a magnetically rolling Janus particle within different geometrical optoelectronic confinements through hybrid-mode propulsion.** (a) Rectangular confinement, where I1–I4 mark the successive positions of the JP and the corresponding orientations of the external magnets. At each stage, the magnets were reoriented to change the rolling direction, driving the JP sequentially along the rectangular confinement path from I1 to I4. (b) Triangular optical confinement, where II1–II3 indicate the successive positions of the JP and the corresponding magnet orientations used to guide it along the trajectory. The applied electric field is 5 MHz, 20 Vpp. Magnetic rotation: 30 RPM.

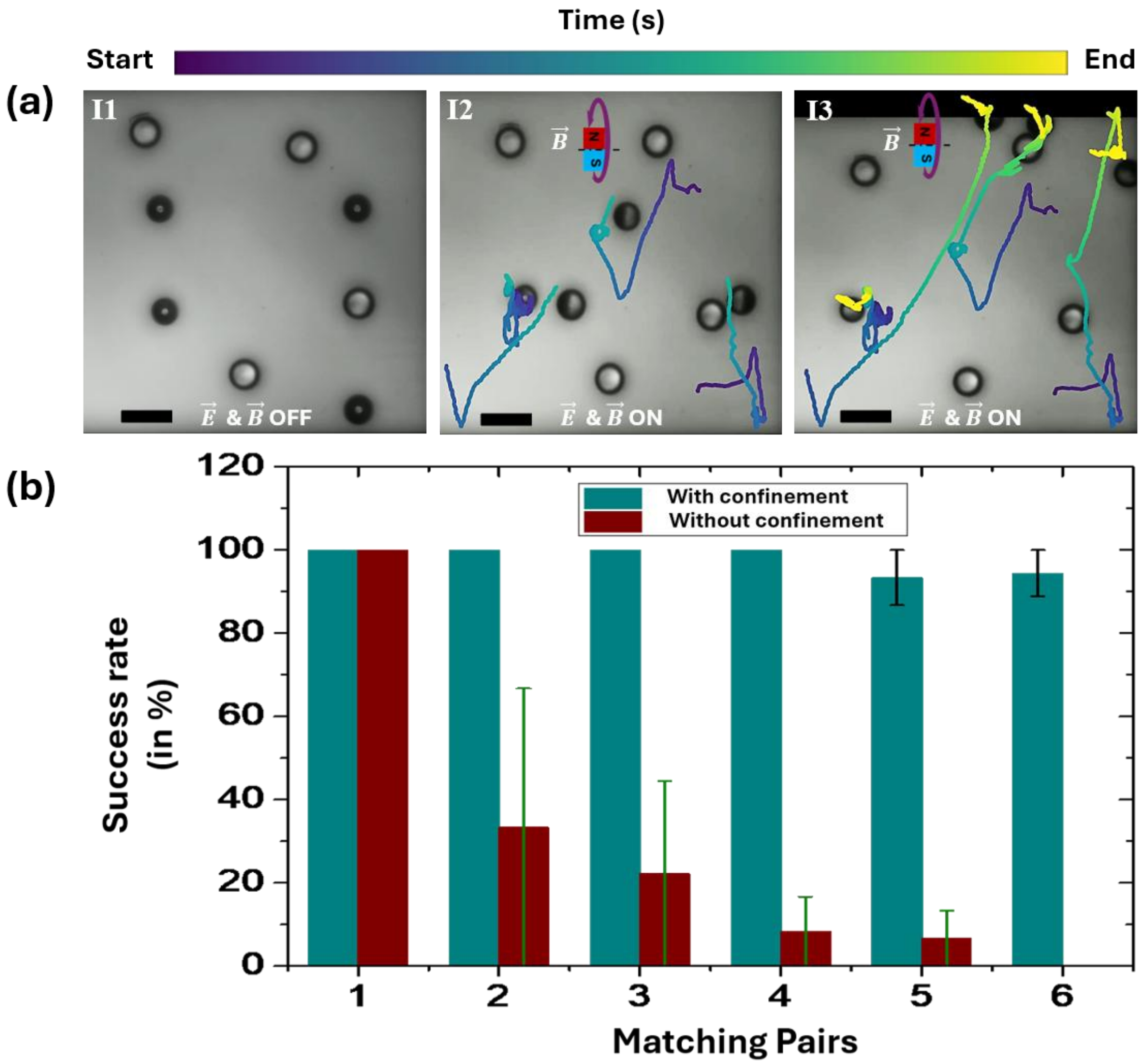

**Figure S9. Pairing under hybrid electric-magnetic field control on the optoelectronic platform.** (a) Images I1-I3 show uncontrolled pair formation under uniform illumination at different time points. Viridis-colored trajectories illustrate the JP motion over 18 s. An electric field of 5 MHz and 20 Vpp and a magnetic rotation rate of 30 RPM in the indicated direction were applied. Scale bar: 50 µm. (b) Pairing success rate under optoelectronic confinement was significantly higher than that under unconfined uniform illumination.

**Movies**

Movie S1: Simultaneous, decoupled motion of active particles (JP and '+' shaped particles) along linear and complex geometrical trajectories, corresponding to Figure 1d-i. An AC electric field of 50 kHz and 15 Vpp was applied.

Movie S2: Manual active JP- passive PS pairing using optoelectronic confinement and a control experiment without confinement, corresponding to Figure 2a-b. An AC electric field of 50 kHz and 15 Vpp was applied.

Movie S3: Simultaneous pairing of JPs with cells through optoelectronic confinement, corresponding to Figure 2d. An AC electric field of 50 kHz and 7 Vpp was applied.

Movie S4: Attachment of a JP to a cell followed by electroporation in an ITO chip, corresponding to Figure S3. An AC electric field of 50 kHz and 7 Vpp was applied for particle approach and attachment, followed by 1 kHz, 20 Vpp pulses for electroporation.

Movie S5: Heuristic simulation of algorithm-based automated pairing and experimental demonstration of multi-agent microrobotic operation, corresponding to Figure 3b and 3c. An AC electric field of 50 kHz and 15 Vpp was applied.

Movie S6: Decoupled motion of JPs under globally applied magnetic field through optoelectronic confinement, corresponding to Figure 4b. An AC field of 5 MHz and 20 Vpp and a magnetic rotation rate of 30 RPM were applied.

Movie S7: The hybrid electric-magnetic propulsion of a JP along rectangular and triangular optical trajectories, corresponding to Figure S8. An AC field of 5 MHz and 20 Vpp and a magnetic rotation rate of 30 RPM were applied.

Movie S8: The pairing formation between JP and PS through hybrid mode of magnetic field-induced propulsion and optoelectronic confinement, corresponding to Figure 4c. An AC field of 5 MHz and 20 Vpp and a magnetic rotation of 30 RPM were applied.